\documentstyle[seceq,epsf]{ptptex}

\title{
Luminous Matter Distribution, Bulk Flows and\\ Baryon Content
in Cosmological Models with a Local Void
}

\author{
Kenji {\sc Tomita}\footnote{E-mail address: tomita@yukawa.kyoto-u.ac.jp
} 
}

\inst{
Yukawa Institute for Theoretical Physics, Kyoto University, \\
Kyoto 606-8502, Japan}

\recdate{February 27, 2002      
}

\abst{
First, we consider galaxy formation from the viewpoint of
hierarchical clustering theory and discuss the possibility that
inhomogeneous models with a local  
void may be compatible with the observed homogeneity of galactic
distributions found in  recent redshift surveys, because their
inhomogeneity can be weakened by the
difference in the feedback system of galaxy formation between the inner and
outer regions. Next, it is shown with the results of numerical
simulations that the observed inhomogeneity of two-point
correlations of galaxies can be accounted for by these models. Also,
the natural appearance of bulk flows for an off-central observer is
demonstrated. Finally, the inhomogeneity of the baryon content is
discussed from the viewpoint of our inhomogeneous models.
}

\begin{document}

\maketitle

\section{Introduction}
Under the assumption of spatial homogeneity, cosmological
parameters have been determined using important observational
results, such as
the [magnitude $m$ - redshift $z$] relation of type Ia 
SN\cite{sch,riessa,riessb,perl} and the anisotropy of cosmic microwave 
background radiation (CMB),\cite{lange,lee,stomp,hal,pry} and 
it has been found that models with a dominant 
cosmological constant best describe the SN data and flat models best
describe the CMB data. 

The present author\cite{tma} has considered models with a local void 
on scales of $\sim 200$ Mpc in order to explain 
the existence of the large-scale bulk flows measured by 
Hudson et al.\cite{hud} and Willick,\cite{will} and has
shown\cite{tmb,tmd,tme} that the zero $\Lambda$ and small $\Lambda$
models with flat space in the 
outer region are consistent with the data of type Ia supernovae 
(especially the recent data for $z = 1.7$\cite{new}) and with the CMB
anisotropy. It thus appears that these models are competent for
accounting for the cosmological
observations in spite of the small realization probability of their
large-scale void. Here the void means a low-density region with respect to
the total matter, which consists mainly of the dark matter.

The homogeneity of galactic distributions found in recent redshift
surveys may support homogeneous models, but it does not necessarily rule out 
the inhomogeneous models with a local void, in which the galactic 
number densities in
the inner and outer regions may be comparable owing to a larger
suppression of galaxy formation in the outer region.

In this paper we consider in \S 2 the process of galaxy formation
and observational aspects of inhomogeneous models from the viewpoint 
of hierarchical galaxy formation theories. To demonstrate the
feasibility of the
inhomogeneous models, it is necessary to show how they can predict a
nearly homogeneous distribution of galaxies in spite of their void
structure. Here we study the observed galactic distributions in
connection with the existence of a local void and  
the two-point correlation that reflects gravitationally the
inhomogeneous structure of our models. In \S 3, we describe the models
with a local void  
using numerical simulations, and in \S 4, we analyze the theoretical 
two-point correlations ($\xi^{\rm I}$ and $\xi^{\rm II}$) in the inner
and outer regions and show that the computed ratio 
$\xi^{\rm I}/\xi^{\rm II}$ is consistent with the observed 
ratio. In \S 5, we derive the bulk flows for  an off-center 
observer.  In \S 6 we discuss the inhomogeneity of the baryon
content from the point of view provided by our models. Section 7 is
dedicated to concluding remarks.   

\section{Galaxy formation and observational aspects}
Modern theories of galaxy formation led to a great deal of progress in 
the understanding of galactic properties under the assumption of cold
dark matter (CDM) in hierarchical clustering cosmologies. In these
models, it is 
assumed that galaxies form when gas cools and condenses in dark
matter halos that merge during their evolution. The key
point in these theories is that the energy released from stars acts as 
a ``negative feedback'' on the gas and star formation. This idea was 
proposed by White and Rees\cite{wr} and developed by White and 
Frenk.\cite{wf} 

Recent works on galaxy formation, including the process of 
feedback, have employed two important methods, a semi-analytic method and 
numerical simulations. The first is directly connected with the
above two pioneering works and takes into account the feedback in 
detail, but has been applied only in simplified (spherically
symmetric) situations.  
The merger history of a dark matter halo has been treated using either a
statistical method (Kauffman et al.,\cite{kauf94} 
Cole et al.,\cite{cole94,cole20} Somerville and Primack\cite{som}) 
or with $N$-body 
simulations (Kaufmann, Nusser and Steinmetz,\cite{kns} Kauffmann,
Colberg, Diaferio and White,\cite{kcdw} 
Benson et al.,\cite{bens} Somerville et al.\cite{slsd}). 
In the case of simulations, both the merger history (using 
$N$-body simulations) and the feedback
process and star formation (using gas dynamical simulations) have been 
investigated (Cen and 
Ostriker,\cite{cen} Katz, Weinberg and Hernquist,\cite{katz} Frenk 
et al.,\cite{frenk} Kay et al.\cite{kay}). 
This method has the advantage that no artificial symmetries need to 
be imposed, but it has the disadvantage that the treatment is not  
transparent and there are unrealistic limits on the dynamic range of
resolved structures, though the resolution is being improved with the 
use of high-speed supercomputers. In the following, we first give an
 outline of recent treatments of galaxy formation on the basis of 
 the semi-analytic method of Kauffmann et al., and next consider how we 
should treat galaxy formation in the present models with a local
void.

\subsection{Galaxy formation in homogeneous models}
The semi-analytic models of Kauffmann et al.\cite{kcdw} consist of a 
combination of $N$-body simulations and an analytic approach 
to star formation and supernova feedback. 
The $N$-body simulations are performed using the cluster normalization
by determining $\sigma_8$, the dispersion of density perturbations, 
within $8 h^{-1}$ Mpc spheres in two CDM models: SCDM ($\tau$model)
and $\Lambda$CDM, with 
$$(\Omega_0, \lambda_0, h, \sigma_8, \Gamma, f_b) 
= (1.0, 0.0, 0.5, 0.6, 0.21, 0.1), \ (0.3, 0.7, 0.7, 0.7, 0.9, 
0.21, 0.15),$$ 
respectively, where $\Gamma$ is the shape parameter of the power 
spectrum and $f_b$ is the baryon factor ($\equiv \Omega_b/\Omega_0$).

The merging history of dark matter halos is constructed using the 
halo catalogues that consist of halos larger than the least stable 
system (containing 10 particles). In each halo, there is a central 
particle, representing the central galaxy onto which gas in a halo falls
and where stars form. The central galaxy in a halo is the central 
galaxy in its most massive progenitor, and the central galaxies in 
less massive progenitors are satellites in the halo. These galaxies 
also grow from small galaxies to larger galaxies through merging.

The physical properties of galaxies are determined by gas cooling, 
star formation, supernova feedback, dust extinction, etc. The most
important processes among them are star formation and supernova feedback. 
Here, a star formation rate of the form
\begin{equation}
  \label{eq:g1}
dM_*/dt \ (\equiv \dot{M}_*) \ = \alpha M_{\rm cold}/t_{\rm dyn}  
\end{equation}
is assumed, where $\alpha$ is a free parameter, $t_{\rm dyn}$ is the
dynamical time of the galaxy, and $M_*$ and $M_{\rm cold}$ are the 
total masses of stars and cold gas in the halo, respectively.

The energy ejected by a supernova explosion into the interstellar 
medium reheats the cold gas to the virial temperature.
The reheating rate is expressed as
\begin{equation}
  \label{eq:g2}
dM_{\rm reheat}/dt = \epsilon {4 \over 3} \dot{M}_* 
\eta_{\rm sn} E_{\rm sn}/{V_c}^2,
\end{equation}
where $\eta_{\rm sn}$ is the number of supernovae per solar mass of
stars ($= 5 \times 10^{-3}/{M_\odot}$), $E_{\rm sn}$ is the kinetic 
energy of the ejection from each supernova ($\cong 10^{51}$ erg), 
$\epsilon$ is a free parameter representing the fraction of this energy 
used to reheat cold gas, and $V_c$ is the circular velocity of galaxies.

The evolution of distributions of dark matter halos and galaxies 
within them is determined using Eqs. (\ref{eq:g1}) and (\ref{eq:g2}), 
when we specify the free parameters $\alpha$ and $\epsilon$, and the 
present distributions can be compared with observations in the 
form of luminosity functions, Tully-Fisher relations and 
two-point correlations.  In order to choose the best values of 
$\alpha$ and $\epsilon$, a normalization condition is imposed: a 
fiducial reference galaxy (which is defined as a central galaxy with 
$V_c = 220$ km s$^{-1}$) is assumed to satisfy the $I$-band Tully-Fisher 
relation. Then the set $(\alpha, \epsilon)$ is given as $(0.07, 0.15)$ 
and $(0.1, 0.03)$ for the SCDM ($\tau$CDM) and $\Lambda$CDM models, 
respectively. These values imply that the SCDM model has smaller star 
formation and larger feedback than the $\Lambda$CDM model. For these
sets, both the SCDM and $\Lambda$CDM models have luminosity functions, 
Tully-Fisher relations and two-point correlations consistent with the 
observed ones. In addition to the feedback due to supernova explosions, 
 photoionization and heating due to ultraviolet radiation also 
contribute to the feedback.\cite{thoul,est2,nags} 

\subsection{Galaxy formation in models with a local void}
In these models the total matter (consisting mainly of dark matter)
has different uniform densities in the inner region ($r < r_b$) and
the outer region ($r > r_b$), and, similarly the luminous matter
(consisting of galaxies) may also have different densities in the two
regions. If the star formation and the feedback effect in the two regions act 
 in a cooperative manner, however, the distributions 
of the luminous matter can be nearly uniform throughout the two regions. 

Let us assume for instance that the outer region is represented by 
the Einstein-de Sitter model and the inner region is represented by an
open model 
with $(\Omega_0, \lambda_0) = (0.3, 0)$. Then we can have $(0.07,
0.15)$ as the set $(\alpha, \epsilon)$ in the 
outer region. Because the model in the inner 
region is similar to the low-density model with $(\Omega_0, 
\lambda_0) = (0.3, 0.7)$, we have \ $(\alpha, \epsilon) \approx (0.1, 
0.03)$\ in the central part of the inner region, say $r \leq 
r_{\rm in} \ (\leq r_b)$. 

In the transient region satisfying $r_{\rm in} \leq r \leq r_b$, the values of 
the set $(\alpha, \epsilon)$ change gradually from the value
$(0.07, 0.15)$ in the outer region to the value $(0.1, 0.03)$ in the
inner region. The strong photoionization by
UV radiation from the outer region to the inner region may play an
effective role in preventing small galaxies from forming and hence growing
through merging in this region. For this reason, the number density of
forming galaxies is small there, compared 
with the densities in the outer and inner regions.

Galaxies formed in the transient region enter the outer region thereafter,
colliding with other galaxies there, since the inner expanding velocities 
are larger than the outer velocities. At the present epoch, the boundary 
$(r = r_{\rm in})$ of the inner central region reaches the boundary 
$r = r_b$ of the outer region, so that the number densities of galaxies 
in the outer and inner regions may be nearly equal, as shown in Fig.~1.
At present, however, we cannot determine the position $r = r_{\rm
in}$ accurately, since studies on the  
dependence of rates of star formation, supernova explosion and  
UV radiation on the background models have not yet been 
established.\cite{efst} Depending on these rates and the position of
the central part ($r < r_{\rm in}$), the various large-scale
structures may appear around 
the boundary ($r = r_b$). It is found in simple simulations 
that, if radial distances for $r_{\rm in} \leq r \leq r_b$ are
about $1/4$ of those satisfying $0 \leq r \leq r_b$ on the comoving scale, the
two regions become smoothly connected later.

\begin{figure}
\epsfxsize = 14cm
\centerline{\epsfbox{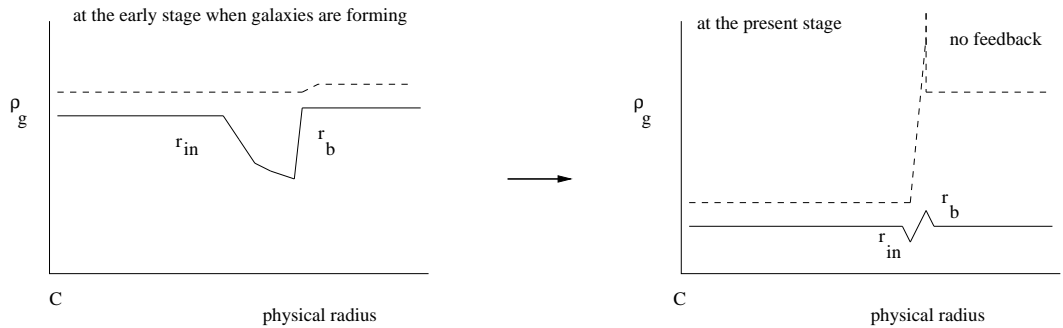}}
\caption{The schematic evolution of galactic density $\rho_g$ around
the boundary $r_b$. The dashed curves correspond to the case of no feedback.
 \label{fig:1}}
\end{figure}

 Here we have the important prediction that around
the boundary there must be many active galaxies  
that are born as a result of the frequent collisions of galaxies,
even if the distribution of galaxies is smooth there owing to
strong feedback.

It is important to investigate the evolution of luminous matter using
numerical simulations, taking into account the above described complicated
processes including cooling, star formation, supernova feedback and
heating by UV radiation, but such investigations are far beyond the
scope of this 
paper. At present, we cannot show quantitatively how luminous 
matter becomes nearly homogeneous, but we are able to describe a
possible process leading to the approximate homogeneity.

\subsection[]{Observed galactic distributions}

Here we consider the present observational situation regarding the galactic
distribution from the viewpoint of our inhomogeneous models.

\subsubsection{Dependence on the observer's position}

Several redshift surveys have suggested the existence of 
a low density region or a local void with radius $\sim 300h$ Mpc
($H_0 = 100h$ km sec$^{-1}$ Mpc$^{-1}$).\cite{mari,marz,folk,zucc}
 The latest surveys (2dF and SDSS)
show that the galactic distributions within a radius $\sim 600h$ Mpc
are homogeneous on the whole, though they include various large-scale
structures. Thus the observed distributions seem to be consistent with
homogeneous cosmological models. However, inhomogeneous models with a
local void also may be compatible with these surveys for the following 
reasons:
 
\noindent (1) The number densities of galaxies in the inner low-density
region and the outer high-density region may be similar in these
models also, as discussed in the previous subsection.  

\noindent (2) Some of observed large-scale structures, which are
regarded as independent, may represent correlated structures near 
the boundary. For an observer in the center, these correlated structures 
have an equal redshift and can easily be distinguished. For us
(off-center observers), however, they have direction-dependent 
redshifts and can be confused with other uncorrelated structures.
When we superpose the redshift histograms in various directions 
to the north and south, off-central observers may see the structures 
there as independent at different distances.

\subsubsection{Active galaxies around the boundary}

From a recent study of spectral types of galaxies in the 2dF
survey (Madgwick et al.\cite{mad}), it was found that there are 
many active galaxies in the redshift region corresponding 
to the boundary $r = r_b$. These galaxies may have been born
as a result of the above-mentioned collisions of galaxies around 
the boundary. Similar analyses of the spectral types of galaxies are
necessary for the SDSS survey to clarify the astrophysical 
situation around the boundary.
 
\subsubsection{Two-point correlation}

The apparent distribution of luminous matter may not be useful in the
study of the
background model consisting mainly of dark matter, because of the very
complicated processes of formation and evolution of galaxies. The
two-point correlations of galaxies, on the other hand, depend mainly
on the background model, at least for less luminous galaxies in the 
homogeneous regions. Therefore it may be important to use these 
correlations to study the distribution of dark matter also. 

Recently, large-scale redshift surveys have provided valuable
information about the spatial dependence of the two-point correlations
of galaxies in the region $z < 0.1$. Using the NGP + SGP data from
the 2dF surveys, Norberg et al.\cite{norberg} derived the correlation 
length $r_0$
for samples with the various parameters ranges (cf. Table~1 and Fig.~3
in their paper). The observed correlation length increases with the
absolute magnitude as \
$r_0 = 4.14 \pm 0.64, \ 4.43 \pm 0.45, \cdots ,
9.38 \pm 1.48$ \
 for \ $(M_{bJ} - 5\log_{10} h, z) = (-18.0$ -- $-18.5, \
0.010$ -- $0.080), \ (-18.5$ -- $-19.0, 0.013$ -- $0.104), \cdots , (-21.5$
-- $-22.5, 0.059$ -- $0.28)$, 
respectively, where $M_{bJ}$ is the
$b$-band absolute magnitude. The SSRS2, EPS and Stromlo
data also exhibit a trend similar to that found in the 2dF survey, as
shown in Fig.~3 in
Norberg et al.'s paper.\cite{norberg} The luminosity dependence of 
$r_0$ in these
samples seems to be consistent with that discussed in the
hierarchical galaxy formation theories. 

According to a recent theory
of galactic distribution (Benson et al.\cite{bens}), $r_0$ is an
increasing function of
\ $- M_{bJ}$ for $-(M_{bJ} - 5\log_{10} h) > 21$,
while it is constant or a slightly decreasing function of \ $- M_{bJ}$ for
$-(M_{bJ} - 5\log_{10} h) < 21$. The observed value of $r_0$ is an
increasing function of $- M_{bJ}$ even for $-(M_{bJ} - 5\log_{10} h) <
21$. Quantitatively, it increases by a factor of about $1.5$ over the
interval $(M_{bJ} - 5\log_{10} h) = -18$ -- $-21$. The two-point
correlation $\xi \propto r_0^\gamma \ (\gamma \approx 1.7)$ changes by
a factor of $2.0$ over this interval. We can interpret this change to 
represent the redshift
dependence (or the spatial inhomogeneity) of $r_0$ or $\xi$ in the
region $z = 0.010$ -- $0.280$. The boundary ($z \sim 0.07$) 
of the local void we consider is included in this region.  

From the data of the SDSS survey, Zehavi et al.\cite{zehavi} derived the
correlation length $r_0$ for three samples, obtaining 
\ $r_0 = 7.42 \pm 0.33, \
6.28 \pm 0.77$ and $4.72 \pm 0.44$ 
 for \ $(M^*_r, z) = (-23.0$ --
$-21.5, 0.100$ -- $0.174), \ (-21.5$ -- $-20.5, 0.052$ -- $0.097),$ and
$(-20.0$ -- $-18.5, 0.027$ -- $0.051)$, 
respectively \ (cf. Table~2 and Fig.~16 in their paper.)      
Their result also reveals a change in $r_0$ by a factor of about
$1.33$ over the interval \ $M^*_r = -21.5$ -- $-18.5.$ This change can
be interpreted as the redshift dependence of $r_0$ in the interval 
$z = 0.027$ -- $0.097$.  Our boundary ($z \sim
0.07$) is also included in this interval.  

In order to explain these spatial changes in $r_0$ and $\xi$, we
study inhomogeneous models with a local void in the following
sections using simple simulations. 

\section{Numerical inhomogeneous models}

In previous papers, I used models with a local void consisting of
homogeneous inner and outer regions with a singular shell or an
intermediate self-similar region. In order to study the evolution of
nonlinear perturbations and their two-point correlations in these
regions, I derived numerical inhomogeneous models
using the method of $N$-body simulations, by considering a spherical
low-density region in the background homogeneous models. These models
are presented in this section. The
background model parameters are expressed as $H_0, a_0, \Omega_0$ and
$\lambda_0$, representing the
Hubble constant, the scale factor, the density parameter and the
cosmological-constant parameter at the present epoch $t = t_0$,
respectively, and 
the spatial curvature is $K = a_0^2 H_0^2 (\Omega_0 +\lambda_0-1)$. 
 The proper radial distance $R$ at $t_0$ is related to the radial 
coordinate $r$ by $R = a_0 r$. As the background models, we consider the
following three cases:
$(\Omega_0, \lambda_0, h)$ = \ (1) Einstein-de Sitter model $(1.0, 0.0,
0.5)$, \ (2) open model $(0.6, 0.0, 0.6)$ and \ (3) flat
nonzero-$\Lambda$ model $(0.6, 0.4, 0.6)$.  
 
The $N$-body simulations were performed using the tree-code used by
Suto and Suginohara, in which a periodic condition is imposed, and the
initial perturbed state is determined using COSMICS. 
Here $N = 2.1 \times 10^6$, and the particle mass $M$ is $M = 5.7, 2.9$ and
$2.9 \times 10^{13} M_\odot$ for the above three cases (1), (2) and
(3), respectively. The softening radius is 1 Mpc for all cases.
The periodic condition is given at $|x^1| = |x^2| = |x^3| = r_p$,
where $R_p \equiv a_0 r_p = 300/h$ Mpc and $r \equiv [(x^1)^2 +
(x^2)^2 + (x^3)^2]^{1/2}$. 

The initial conditions are
set at $t = t_i$ at which $z = z_i \ (=15.5)$, in the form of displacements
$\delta x^k$ and velocities $v^k$ of $N$ particles ($k = 1, 2, 3$). 
After the initial conditions in the homogeneous case are given, a
low-density region is introduced at the initial epoch by changing 
the particle positions $x^k$ in the inner region ($r < r_b$) and the
intermediate region ($r_b < r < r_{b1}$) as 
\begin{equation}
  \label{eq:m2}
x^k = (x^k)_{\rm hom} \times  \left\{ 
\begin{array}{c}
(1 + d) \hspace{3.3cm} {\rm for}\quad r < r_b, \cr 
 \Big[1+ d\Big({1 \over r_b} - {1
\over r_{b1}}\Big)/\Big({1 \over r} - {1 \over r_{b1}}\Big)\Big] \quad
{\rm for} \quad r_b \leq r < r_{b1},
\end{array}
\right.
\end{equation}
and $x^k = (x^k)_{\rm hom}$ for $r > r_{b1}$ \ (see Fig.~2). Here
$(x^k)_{\rm hom}$ represents the particle positions in the homogeneous case
and $d$ is a constant expansion factor adjusted so as to give the 
expected density parameter and average expansion rate in the inner
region. The intermediate region was introduced to make smooth the
change in the density and velocity of particles near the boundary.
Here, we treat mainly the case \ $R_b \equiv a_0 r_b = 180/h$ Mpc
and $R_{b1} \equiv a_0 r_{b1} = 210/h$ Mpc, and  consider also the
case \ $R_b = 120/h$ Mpc and $R_{b1}= 140/h$ Mpc for comparison.  

\begin{figure}
\epsfxsize = 9cm
\centerline{\epsfbox{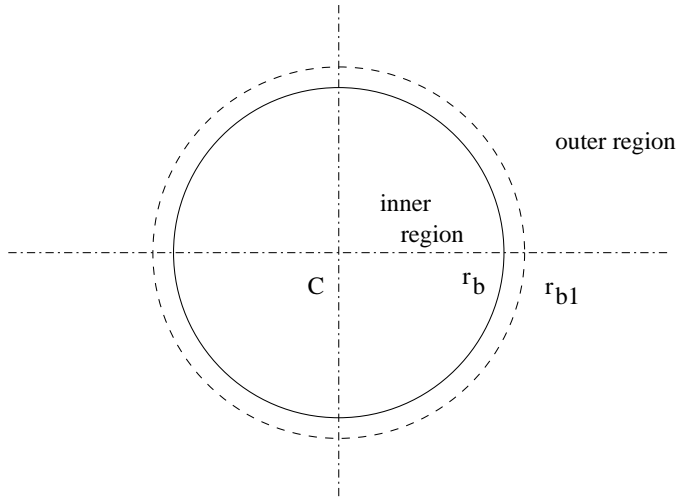}}
\caption{Two regions and the boundary. The solid and dotted curves denote
the surfaces with $r = r_b$ and $r = r_{b1}$, respectively.\label{fig:2}}
\end{figure}

From the viewpoint of our models with a local void, the above model 
parameters describe the outer region ($r > r_b$) and correspond to 
$({\Omega_0}^{\rm II}, {\lambda_0}^{\rm
II}, h^{\rm II})$ in the previous papers,\cite{tma,tmb} where
$H_0^{\rm II} = 100 h^{\rm II}$ km s$^{-1}$ Mpc$^{-1}$.
From the simulations, the parameters  $({\Omega_0}^{\rm I},
{\lambda_0}^{\rm I}, h^{\rm I})$ in the inner region are derived as
their average values in the expression
\begin{equation}
  \label{eq:m3}
H_0^{\rm I} \equiv H_0^{\rm II} + \Big(\sum_{i \in I_*}
v_{(i)}/r_{(i)} \Big)/N_*
\end{equation}
along with $\lambda_0^{\rm I} \equiv {1 \over 3} \Lambda/(H_0^{\rm I})^2$,
where the summation is taken over the set $I_*$ of particles included in the
region\ $r_l < r < r_b \ (r_l \sim 0.2 r_b)$, so 
as to avoid the disturbances near the origin and the boundary. Here, $N_*$
is the particle number in this region, and $v_{(i)}$ and $r_{(i)}$
are the radial velocities and radii of the $i$-th particle, where $v_{(i)}
\equiv (\sum_{k=1}^3 v_{(i)}^k x_{(i)}^k)/r_{(i)}$.

For the particles that were initially in the inner region, the
observed positions (in redshift space) are different from those
represented by the background coordinates (in real space). To take
the average velocity (in the inner region) into account, we define
other coordinates (average comoving coordinates) ${\bar{x}}^k \ 
(k = 1, 2, 3)$ as
\begin{equation}
  \label{eq:m4}
{\bar{x}}^k = x^k \times  \left\{ 
\begin{array}{c}
H_0^{\rm I}/H_0^{\rm II} \hspace{4.6cm} {\rm for}\quad r < r_b, \cr 
 \Big[1+ \Big({H_0^{\rm I}\over H_0^{\rm II}} - 1\Big)\Big({1 \over r_b} - {1
\over r_{b1}}\Big)/\Big({1 \over r} - {1 \over r_{b1}}\Big)\Big] \quad
{\rm for} \quad r_b \leq r < r_{b1},
\end{array}
\right.
\end{equation}
and ${\bar{x}}^k = x^k$ for $r > r_{b1}$. 
This coordinate system constitutes both an outer comoving system and
an inner comoving system with respect to the mean 
motion. 
Using the volume $V_* = {4 \over 3} \pi a_0^3 (r_b^3 - r_l^3) 
(H_0^{\rm I}/H_0^{\rm II})^3$  in these coordinates, the density
parameter in the inner region is defined as
\begin{equation}
  \label{eq:m5}
\Omega_0^{\rm I} \equiv {3\pi G \over 3 (H_0^{\rm I})^2} {M N_*  \over V_*}.
\end{equation}
These values in the above three cases are listed in Table~I together
with $d$ and $R_b$. The values of $d$ were chosen so as to obtain
$\Omega_0^{\rm I} \sim 0.3$ for three sets of $(\Omega_0^{\rm II},
\lambda_0^{\rm II})$. By comparing the case $R_b h = 180$ and the
case $R_b h = 120$, we find that the choice of the position of the
boundary is not crucial.

\begin{table}
\centering
\caption{Inner model parameter values determined statistically for given 
outer model parameters  $(\Omega_0, \lambda_0, h) = ({\Omega_0}^{\rm II}, 
{\lambda_0}^{\rm II}, h^{\rm II})$.  Here $d$ is a constant expansion
factor and $R_b$ is the radial distance between the center C and the
boundary. $\Omega_b$ is the present baryon density parameter ($\propto
\Omega_0$). } 
\label{tab:1}
\begin{tabular}{|c|c|c|c|c|} \hline
{${\Omega_0}^{\rm II} \ {\lambda_0}^{\rm II} \ h^{\rm II}$} & {$d$}
&{$R_b h$ Mpc} & {${\Omega_0}^{\rm I}\ \ {\lambda_0}^{\rm I}\ \ h^{\rm
I}$} & $\Omega_b^{\rm II} (h^{\rm II})^2/\Omega_b^{\rm I} (h^{\rm
I})^2$ \\ \hline  
$1.0\ \ 0.0\ \ 0.50$ & $0.022$ & $180$ & $0.38\ \ 0.00\ \ 0.576$ & $1.98$ \\ 
$1.0\ \ 0.0\ \ 0.50$ & $0.022$ & $120$ & $0.39\ \ 0.00\ \ 0.574$ & $1.98$ \\
$1.0\ \ 0.0\ \ 0.50$ & $0.030$ & $180$ & $0.32\ \ 0.00\ \ 0.594$ & $2.21$ \\
$0.6\ \ 0.0\ \ 0.60$ & $0.040$ & $180$ & $0.29\ \ 0.00\ \ 0.693$ & $1.55$ \\
$0.6\ \ 0.4\ \ 0.60$ & $0.040$ & $180$ & $0.29\ \ 0.29\ \ 0.694$ & $1.55$ \\
  \hline
\end{tabular}
\end{table}

The distribution of dark matter particles at present epoch is displayed
in Fig.~3 as
an example for the model parameter set $(1.0, 0.0, 0.5),$ where the
particles in the range \ $-3\ {\rm Mpc}\leq a_0 x^3 \leq 3\ {\rm Mpc}$
are plotted. 
The corresponding distributions of
particles in the $\bar{x}^k$ coordinates are displayed in Fig.~4.
It is found that at the present epoch, the (dark matter) particles in the
inner region near the boundary seem to have been
mixed with those in the outer region, so that their
distributions are quite complicated in the outer region near the boundary.  

\begin{figure}
\epsfxsize = 9cm
\centerline{\epsfbox{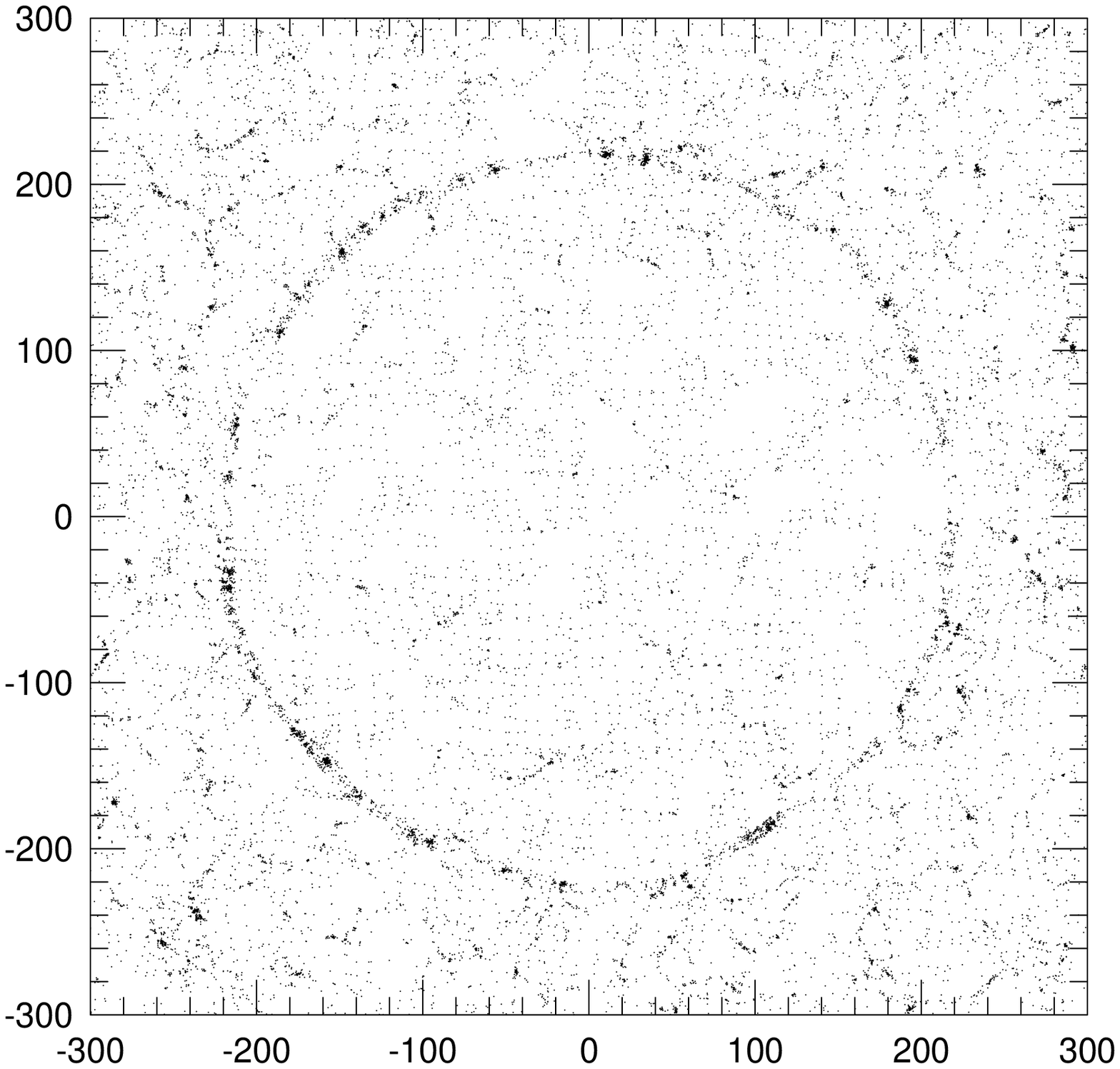}}
\caption{The distribution of dark matter particles in the $(a_0 x^1,
a_0 x^2)$ plane at the present epoch in the case $(\Omega_0^{\rm II},
\lambda_0^{\rm II}) = (1, 0)$ and $R_b h = 180$ Mpc. The length of the 
region displayed is $600/h$ Mpc, and its width is $6/h$ Mpc.    \label{fig:3}}
\end{figure}
\begin{figure}
\epsfxsize = 9cm
\centerline{\epsfbox{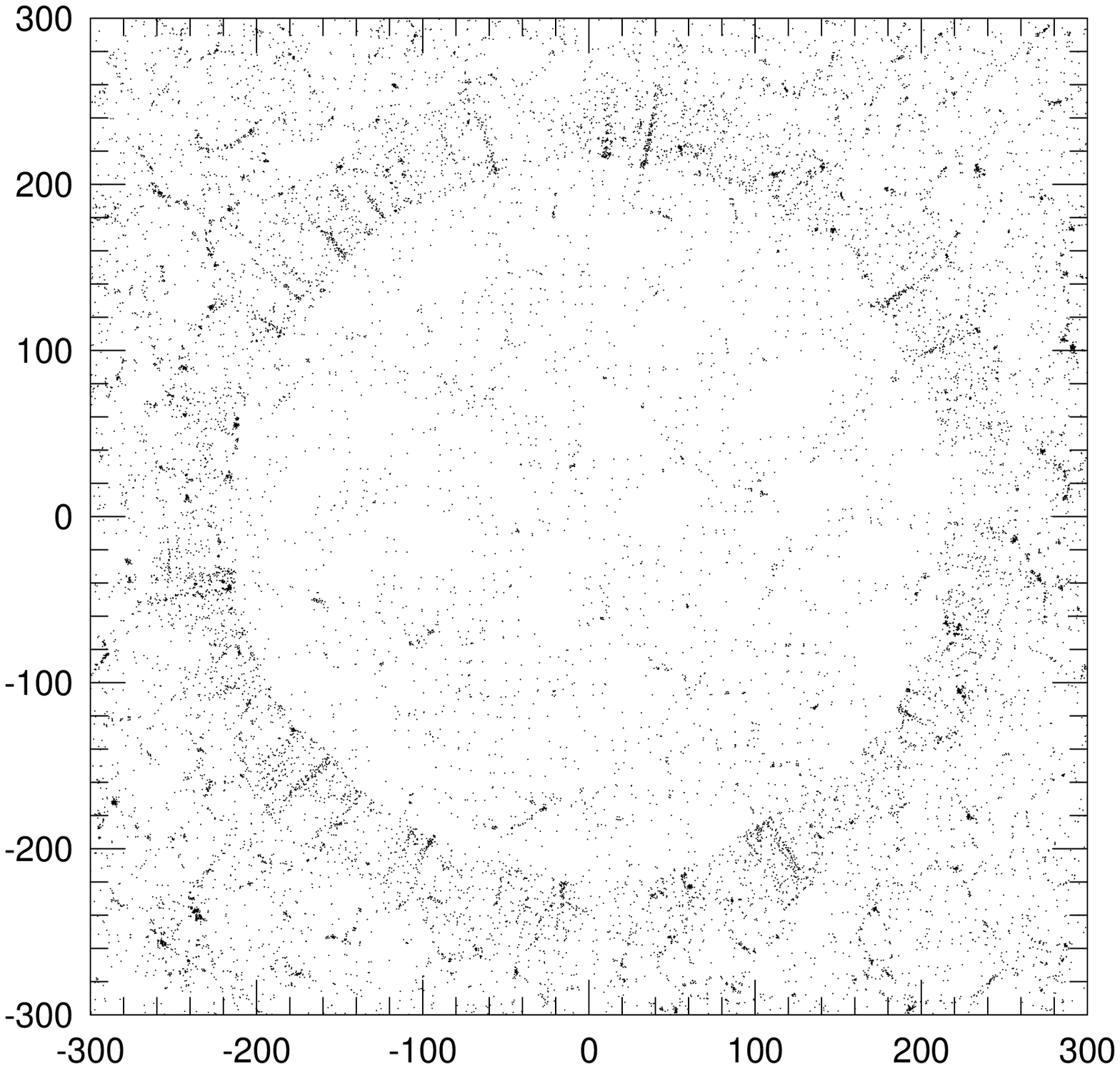}}
\caption{The distribution of dark matter particles in the $(a_0 \bar{x}^1,
a_0 \bar{x}^2)$ plane at the present epoch in the case $(\Omega_0^{\rm II},
\lambda_0^{\rm II}) = (1, 0)$ and $R_b h = 180$ Mpc.  The length of the 
region displayed is $600/h$ Mpc, and is width is $6/h$ Mpc.  \label{fig:4}}
\end{figure}
%

\section{Theoretical two-point correlations}
In this section the two-point correlations ($\xi$) are derived from the
numerical models given in the previous section. Their values in the
inner region, with $r < r_b$, and the outer region, with $r_p > r >> 
r_{b1}$,  have simple behavior, but they are 
complicated in the intermediate region, $r_b < r < r_{b1}$, 
because uncorrelated
particles are mixed there. Here the correlations ($\xi_i$ and $\xi_o$)
in the inner and outer regions, respectively, were calculated using the
particle positions in the average comoving coordinates ($\bar{x}^k$) 
for various cases with $R_b h = 180$ Mpc, which were treated in the 
previous sections. They are assumed to take the form $\xi =
(r_0/R)^\gamma$ with $\gamma = 1.7$ to derive the 
correlation lengths $r_0$. The results are given in Table~II.

\begin{table}
\centering
\caption{Correlation lengths and ratios of two-point correlations
in the two regions of the models with $R_b h = 180$ Mpc. Here, $h =
h^{\rm II}$} 
\label{tab:2}
\begin{tabular}{|c|c|c|c|c|} \hline
{${\Omega_0}^{\rm II} \ {\lambda_0}^{\rm II} \ h^{\rm I}/h^{\rm II}$} 
& {$r_0^{\rm II} h$}&{$r_0^{\rm I} h$} & {$\xi^{\rm I}/\xi^{\rm II}$} 
& $\Omega_0^{\rm II} h^{\rm II}/(\Omega_0^{\rm I} h^{\rm
I})$\\ \hline 
$1.0\ \ 0.0\ \ 1.152$ & $5.2$ & $4.2$ & $1.5$ & $2.3$\\ 
$1.0\ \ 0.0\ \ 1.188$ & $5.6$ & $3.8$ & $2.1$ & $2.6$\\
$0.6\ \ 0.0\ \ 1.155$ & $6.5$ & $5.0$ & $1.6$ & $1.8$\\
$0.6\ \ 0.4\ \ 1.156$ & $5.4$ & $3.9$ & $1.8$ & $1.8$\\
  \hline
\end{tabular}
\end{table}

It is found from this result that the ratio $\xi_o/\xi_i$ is about 2.0
in the case \ ${\Omega_0}^{\rm II} = 1.0,\ {\lambda_0}^{\rm II} =
0.0,\ h^{\rm I}/h^{\rm II} =1.188$, and in other cases, it is
somewhat smaller. Thus the situation in which the ratio of correlations
for dark matter particles is larger than $1.5$ \
is common to these models with a local void.  Here we compare 
this ratio with the observed ratio for galaxies. In general this is
difficult, because the
clustering of galaxies depends not only on the background model, but
also on their luminosity through the bias effect. 
For galaxies with comparatively low luminosity, however, the bias
effect on the clustering seems to be small, and the ratio for dark
matter may be equal to that for galaxies with $M_{bJ} = -18$ --
$-21$. Thus it can be concluded that the theoretical ratio
($\xi_o/\xi_i$) can account for the change in the observed ratio for galaxies
over the range $M_{bJ} = -18$ -- $-21$. 

Now let us consider the power spectra in the two regions. If the
spatial scales are much less than $r_b$ or much larger than $r_b$, 
we can treat the power
spectrum in each region as that of a homogeneous model with the fitting 
formula ($i = $ I and II)
\begin{equation}
  \label{eq:t1}
P_i(k) \propto k \Big[{\ln (1+2.34 q_i) \over 2.34 q_i} \big]^2 [1
+ 3.89q_i + (16.1q_i)^2 + (5.46q_i)^3 + (6.71q_i)^4]^{-1/2},
\end{equation}
where $q_i = k (h^i)^{-1}/\Gamma_i$, and the shape parameter is $\Gamma_i 
\equiv \Omega_0^i  h^i \exp [-\Omega_B^i (1 + \sqrt{2h^i/\Omega_0^i})]$,
depending on the baryon density.\cite{sugi}
In the outer region we can derive $\Omega_0^{\rm II} h^{\rm II}$
from the measurements of CMB anisotropies, but at present we have only 
the rough value $\Omega_0^{\rm II} h^{\rm II} = 0.3$ -- $0.5$, which is
obtained as those with weak prior from Boomerang, DASI and MAXIMA
experiments.\cite{lange,stomp,pry} In the inner region also we can roughly
determine $\Omega_0^{\rm I} h^{\rm I}$ from the treatment of Peacock
and Dodds,\cite{pea} in which 
they derived the reconstructed linear data for the density contrasts of
galaxies and clusters. Their data for the case $k/h < 0.02$ Mpc$^{-1}$
correspond to our outer region and have comparatively large
uncertainty. Therefore, by assigning larger weights to the data for 
$k/h > 0.03$ Mpc$^{-1}$, we
estimate $\Omega_0^{\rm I} h^{\rm I}$ as  $\Omega_0^{\rm I} h^{\rm I} 
= 0.2$ - $0.25$.
Thus we obtain the observational ratio  $\Omega_0^{\rm II} h^{\rm
II}/(\Omega_0^{\rm I} h^{\rm I}) = 1.0$ -- $2.5$ at present.

In our models, by comparison, we have the ratios listed in the 
last column of Table II in the cases (1), (2) and (3). It is seen that
our models correspond to the case with comparatively large ratios.

\section{Bulk flows in the inner region}
If we observe from the center C, the average velocity field would be
 isotropic, and the directions of the velocities of particles would be
 radial at each point in the inner region. For an off-center observer O (being
 a realistic observer), however, the average velocity field is  
anisotropic, and the non-radial component is found to be significant, when 
each velocity($\vec{v}$) in the inner region is divided into 
the component ($v_r$) in the radial direction O $\rightarrow$ A and
the component ($v_p$) in the direction C $\rightarrow$ O as 
\begin{equation}
  \label{eq:b1}
\vec{v} = v_r\ \vec{l} + v_p\ \vec{n}.
\end{equation}
Here  $\vec{l}$ and $\vec{n}$ are the unit vectors in
the directions of O $\rightarrow$ A and C $\rightarrow$ O,
respectively (see Fig.~5), and we have 
\begin{equation}
  \label{eq:b2}
v_r = \Big[\vec{v}\vec{l} -
(\vec{v}\vec{n})(\vec{l}\vec{n})\Big]/\Big[1 - (\vec{l}\vec{n})^2\Big]
\end{equation}
and 
\begin{equation}
  \label{eq:b3}
v_p = \Big[\vec{v}\vec{n} -
(\vec{v}\vec{l})(\vec{n}\vec{l})\Big]/\Big[1 - (\vec{l}\vec{n})^2\Big].
\end{equation}
Then the mean physical value $V_p$ for $v_p$  is defined by
\begin{equation}
  \label{eq:b4}
V_p \equiv a_0 \Big(\sum_{i \in I_*} v_{p(i)} \Big)/N_*,
\end{equation}
where $i$ is the particle index and $N_*$ is the particle number in
the set $I_*$. 
This value depends on the model parameters and is proportional to 
the distance $R_{co} (\equiv a_0 r_{co})$ from
C to O. By choosing O in $N_0$ different directions with equal
$R_{co}$, we can derive the $N_0$ independent values of
$V_p$. Using them, we obtained an average value of $V_p$ 
and the dispersion $(\sigma_p)$ for $N_0 = 6$. Their values are listed
in Table~III for models with $R_b = 180/h$ Mpc and $R_{co} = 50$ or
$60/h$ Mpc, where $h = h^{\rm II}$. The values for $V_{\rm bulk} \equiv
(H_0^{\rm I} - H_0^{\rm II}) R_{co} = 100 (h^{\rm I} - h^{\rm II})
R_{co}$ also are listed, and it is found that $V_p$ is equal to
$V_{\rm bulk}$. In the inner region, the average velocity field of
galaxies is equal to that of dark matter, and it therefore appears that
this velocity corresponds to the bulk flow of clusters 
with the velocity $\sim 700$km s$^{-1}$, which was measured by Hudson
et al.\cite{hud} and Willick.\cite{will} 
 If the measured velocity is smaller,
$V_{\rm bulk}$ can be adjusted to it by taking smaller $R_{co}$,
because $V_{\rm bulk}$ is proportional to $R_{co}$.

\begin{table}
\centering
\caption{Bulk velocities and their dispersions}
\label{tab:3}
\begin{tabular}{|c|c|c|c|c|} \hline
{${\Omega_0}^{\rm II} \ {\lambda_0}^{\rm II}$} &{$ R_{co} h$ Mpc} 
& {$V_p$ km/s}&{$\sigma_p$ km/s} & {$V_{\rm bulk}$ km/s} \\ \hline 
$1.0\ \ 0.0$ &$ 50$ & $764.5$ & $49.5$ & $760$ \\ 
$1.0\ \ 0.0$ &$ 40$ & $752.8$ & $46.6$ & $752$ \\
$0.6\ \ 0.0$ &$ 50$ & $774.5$ & $66.8$ & $775$ \\
$0.6\ \ 0.4$ &$ 50$ & $777.6$ & $57.1$ & $780$ \\
  \hline
\end{tabular}
\end{table}
\begin{figure}
\epsfxsize = 9cm
\centerline{\epsfbox{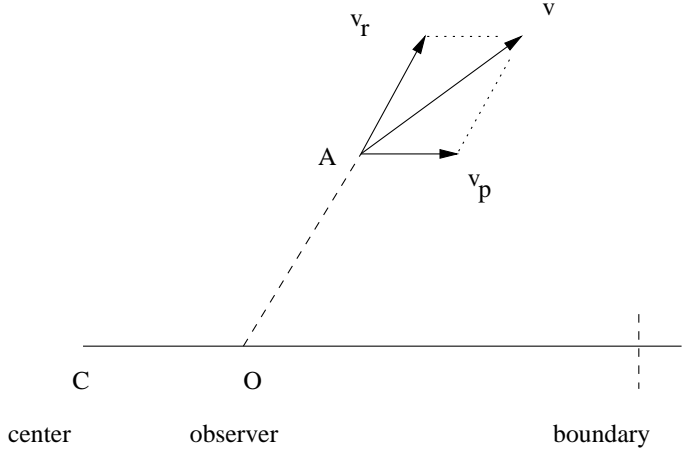}}
\caption{Velocity vectors in the inner region. \label{fig:5}}
\end{figure}
%

\section{Baryon content}

Let us now consider the baryon content in our models with a void. 
Here we assume that all inhomogeneities evolved
gravitationally from perturbations with very small amplitudes at the
stage of nuclear synthesis and that dark matter and baryons are
well mixed, so that the ratio of the baryon density to the matter
density is equal everywhere. Then in the case \   
$(\Omega_0^{\rm I}, \Omega_0^{\rm II}) = (0.3, 1.0)$ and $h^{\rm
II} = h^{\rm I} \times 0.8$ in the inner (I) and outer (II) regions, 
we have $\Omega_0^{\rm II}/\Omega_0^{\rm I} = 1.0/0.3$, so that 
$\Omega_0^{\rm II} (h^{\rm II})^2 = 2.1 \times \Omega_0^{\rm I}
(h^{\rm I})^2$ \  and \ $\Omega_b^{\rm II} (h^{\rm II})^2 = 2.1 \times
\Omega_b^{\rm I} (h^{\rm I})^2$. 
The ratios  $\Omega_b^{\rm II} (h^{\rm II})^2 /\Omega_b^{\rm I} (h^{\rm
I})^2$ for the models we used in our simulations are given in the 
last column of
Table~I. It is found from this table that in models with $\Omega_0^{\rm
II} = 1.0$, the ratio can be about 2, and in models $\Omega_0^{\rm II} =
0.6$, the ratios are somewhat smaller.

From observational studies of light elements,
 it has been found that in the remote past, we have 
$\Omega_b h^2 \cong 0.025$, 
which can be derived from the analysis of deuterium abundance in QSOs 
for $z \sim 2$.\cite{omear,pett,burl} Moreover, recent studies of the CMB
anisotropy suggest similar baryon content, $\Omega_b h^2 =
0.02$ -- $0.03$.\cite{lange,stomp}  
Using these observational results in the outer region and the above
 ratios in the two regions, we obtain $\Omega_b h^2 = 0.010$ -- $0.015$ in the
inner region.  As long as they are produced
adiabatically, the baryon/photon ratio ($\eta$) is equal and constant
($\simeq 7 \times 10^{-10}$) everywhere, and therefore, at the  
stage of primordial nucleosynthesis, the photon number densities also
 were inhomogeneous. At the 
 present stage, however, they have been
equalized in both regions owing to the free propagation of photons.

If the so-called {\it crisis} of big-bang nucleosynthesis\cite{os,steig} 
is actual and $\eta$ is really inhomogeneous (i.e. $\eta \simeq 3 \times
10^{-10}$ in the inner region) in connection with the primordial
abundance of Li$^7$,\cite{olive,ryan} it may be necessary in order to
avoid this {\it crisis}, to invoke that the void was produced as a
special large-scale structure including non-adiabatic perturbations.

\section{Concluding remarks}
In this paper, we first considered galaxy formation in inhomogeneous 
models with a local void. The observed distribution of galaxies 
seems to be homogeneous, but we discussed the possibility that 
this homogeneity may be due to the complicated feedback processes of
galaxy formation and evolution in an inhomogeneous distribution of
dark matter halos. In order to clarify observationally the existence 
of the boundary of the void, it is therefore necessary to investigate 
galactic distributions in greater detail, such as the distribution of active
(colliding) galaxies and two-point correlations.

Next, we derived numerical inhomogeneous models consisting of
dark-matter particles with a spherical low-density (inner) region and
studied the two-point correlations in the inner and outer 
regions. It was found that they can be made consistent with the redshift
dependence of observed correlations of low-luminosity
galaxies. Moreover we derived the bulk flow 
found by an off-center observer, which may correspond to observed
large-scale bulk flow.  Finally, we studied the baryon content and
discussed the possibility of avoiding the {\it crisis} of light elements. 

In previous papers\cite{tmb,tmd,tme} we calculated the luminosity and
angular-diameter 
distances to derive the [$m, z$] relation, and found that the distances in 
our models with a local void are similar to those in
$\Lambda$-dominant homogeneous models in the redshift interval $0 < z<
z_m (\approx 1)$. Thus in the observations using mainly the distances,
such as the number-counts of galaxies and clusters, the theoretical
relations $N(m)$ in the above two different types of models are
similar in the interval $0 < z< z_m$. Even if the observations of
$N(m)$ rule out the SCDM model ($\Lambda =  0$),\cite{yosh,tot} 
our models with a local void and $\Lambda = 0$ may be consistent with the 
observations, like the $\Lambda$-dominant homogeneous models.

It is well known that we are in the low-density region
($\Omega_0 \sim  0.3$), but the total matter density and baryon content in
the remote region ($z > 1$) have not yet been established
clearly through observation. From the observations of the CMB 
anisotropies, we can constrain the model parameters and baryon
content. The recent results of Boomerang, DASI and MAXIMA 
experiments,\cite{lange,stomp,pry} however, seem to differ
and to be partially inconsistent with 
respect to the density parameter and the baryon content. Therefore a precise
measurement by the MAP satellite is needed to
determine these values more accurately. If the additional data from SNIa
with $z > 1.5$  are also obtained from the SNAP 
satellite and large telescopes, like Subaru etc., the model parameters
in the remote region could be determined with much better precision.

\section*{Acknowledgments}
The author is grateful to Y.~Suto and T.~Suginohara for providing 
him with their useful tree code for simulations.
 This work was supported by a Grant-in Aid for Scientific Research 
(No.~12440063) from the Ministry of Education, Science, Sports and
Culture, Japan. He also acknowledges use of the YITP computer system for the
 numerical analyses.


\end{document}